\documentclass{optica-article}

\journal{opticajournal} 

\articletype{Research Article}

\usepackage{lineno}

\begin{document}

\title{Scalable fabrication of  erbium-doped high-Q silica microtoroid resonators via sol-gel coating}

\author{Riku Imamura,\authormark{1} Shun Fujii,\authormark{2} Keigo Nagashima,\authormark{1} and Takasumi Tanabe\authormark{1,*}}

\address{\authormark{1}Department of Electronics and Electrical Engineering, Faculty of Science and Technology, Keio University, 3-14-1, Hiyoshi Kohoku-ku, Yokohama-shi, 223-8522, Japan\\
\authormark{2}Department of Physics, Faculty of Science and Technology, Keio University, 3-14-1, Hiyoshi Kohoku-ku, Yokohama-shi, 223-8522, Japan}

\email{\authormark{*}takasumi@elec.keio.ac.jp} 


\begin{abstract*} 
This study explores sol-gel methods for fabricating erbium-doped silica microtoroid resonators, addressing the limitations of conventional doping techniques and enhancing device scalability. We develop a reproducible sol-gel process that yields defect-free films for photonic applications, and detail common defects and troubleshooting strategies. Two fabrication methods are compared: traditional film deposition on substrates and the direct coating of prefabricated resonators. The latter enables the fabrication of larger resonator diameters (up to 450~\textmu m) without buckling, while achieving a high-$Q$ factor and a low lasing threshold of 350~\textmu W. These erbium-doped resonators exhibit multi-mode laser oscillations at 1550~nm, revealing the sol-gel method’s potential for realizing scalable, gain-doped photonic devices.

\end{abstract*}

\section{Introduction}
In recent years, on-chip photonic integrated devices with optical gain and utilizing rare-earth ions and semiconductor materials have attracted considerable attention~\cite{Polman1997-va,Bradley2011-rg,Mu2020-nf, Xin2019-sq}. Of these, waveguide amplifiers doped with erbium ions have become particularly notable due to their compact size and high efficiency~\cite{Liu2022-ld,Liu2024-qx}, which has brought gain-doped photonic integrated circuits closer to practical realization. Low-loss microcavities with optical gain also offer chip-scale passive mode-locked lasers~\cite{Prugger_Suzuki2021-wi} by incorporating nanomaterials~\cite{Kumagai2018-eg,Tan2021-ib,Fujii2024-bi} that work as saturable absorbers. Another example of photonic integrated devices with gain utilizes non-Hermitian physics~\cite{Wang2023-bn,Chen2018-lk,Zhang2019-lt,Wang2021-tx,Zhang2018-js,Chen2017-sk}, particularly phenomena such as exceptional points. Recent studies have demonstrated that non-Hermitian systems enable us to realize enhanced functionalities, such as improved control over light propagation~\cite{Huang2017-bh}, loss management~\cite{Peng2014-pc,Peng2014-ry}, and enhanced sensor sensitivity~\cite{Zhang2018-js,Chen2017-sk}. Specifically, in coupled whispering-gallery-mode (WGM) resonator systems, a gain-doped resonator is often paired with a passive resonator~\cite{Peng2014-ry}. With this configuration, there is a contrast between gain and loss that depends on the light propagation direction. This has led to several intriguing studies including research on controlling the laser emission direction, suppressing optical modes, and achieving passive mode-locking~\cite{Imamura2024-pa}.

Incorporating ions into existing photonic structures is typically achieved through ion implantation~\cite{Liu2022-ld}, a process in which ions are accelerated under an electric field and directed into a substrate material. This technique allows the precise doping of the material to modify its optical properties. However, a significant drawback of ion implantation is that it can increase losses in waveguides and resonators, as the process may introduce defects and damage to the crystal lattice of the host material~\cite{Min2004-zk}. Annealing is generally required after ion implantation to repair these defects; however, it also presents challenges, such as ion diffusion and potential thermal damage due to internal stress.

In contrast, the sol-gel process~\cite{Orignac1999-km,Zhang2013-sf,Yang2003-zt,Yang2005-ah,Liu2020-zp} is a method for fabricating oxide solids by dissolving organic or inorganic metal compounds in a solution, facilitating hydrolysis and polycondensation reactions within this solution, and further inducing solidification through heating. This technique is predominantly utilized in producing fibers such as silica-alumina, quartz, and alumina fibers in applications involving reflective coatings and other types of coating films~\cite{Liu2020-zp,Huang2002-jq}. This method is initiated with a solution and progresses through a colloidal state containing fine particles, known as “sol,” and transitioning through a “gel” phase with both liquid and gaseous inclusions between solid molecules. This process ultimately allows for the creation of glass or ceramic materials.

This paper has two primary objectives. First, sol-gel methods have been widely adopted for fabricating active nanophotonic devices on-chip, including the development of continuous wave lasing in Er-doped silica toroid microresonators. Although many research groups have independently developed sol-gel processes, the specifics of these techniques are often not fully disclosed, making it challenging for newcomers to the field. In particular, it can be valuable to understand the process parameters that lead to failures, such as membrane cracking, de-wetting, or peeling. A complete, accessible manual would simplify the process, and one objective of this paper is to provide such a guide for the photonics community. Given that the requirements for sol-gel processing in nanophotonics differ from those in other fields, we aim to present a clear, detailed manual specifically tailored for photonics researchers.

Second, this paper reports on two methods for doping on-chip silica toroidal resonators with erbium ions. The first is a conventional approach, where sol-gel silica thin films are deposited on silicon wafers, and resonators are subsequently fabricated from these films~\cite{Yang2005-ah}. Due to the challenges faced when producing very thick films~\cite{Kozuka2004-kp,Yahata2009-td}, this method can encounter issues as regards fabricating large silica microtoroids since buckling may occur[Buckling]. To address these issues, we present a second, novel method in which a sol-gel silica thin film is deposited on a prefabricated resonator structure. The aim when coating an already fabricated resonator is to reduce the effects of buckling while maintaining a high $Q$ factor, providing greater device size flexibility.

These fabrication methods are more straightforward than ion implantation and can be applied to various photonic integrated devices. Their simplicity and versatility make them promising techniques for advancing the development of optical components with enhanced performance.

The paper is organized as follows. Section 2 describes the sol-gel process and details the types of structural failures associated with specific process parameters. Section 3 covers the fabrication of a silica toroid, and Section 4 discusses the coating of onto pre-formed silica toroid microcavities with sol-gel glass. Finally, in Section 5, we provide our conclusions.

\section{Sol-gel thin film process and troubleshooting strategies}
\subsection{Sol-gel thin film fabrication process}
This section provides an overview of the method for fabricating sol-gel silica films on silicon substrates. It explains the defects that can occur at various stages of the film formation process. Fabricating sol-gel thin films involves five major steps: solution mixing, silicon substrate cleaning, spin coating, short annealing, and long annealing. The spin coating and short annealing steps can be repeated for multilayer films to allow us to stack multiple layers.

Previous studies have reported various parameters that allow the successful fabrication of thin glass films using the sol-gel method for optical devices (Table~\ref{table1}). However, because temperature and humidity can significantly impact the film deposition process, these parameters may not directly apply to all experimental environments. Therefore, it is crucial to understand how inappropriate parameters can lead to the appearance of specific failure defects. This study emphasizes fundamental fabrication methods and examines potential defects that may arise in the films due to improper process parameters.

\begin{figure}[htbp]
\centering
\includegraphics[width=\linewidth]{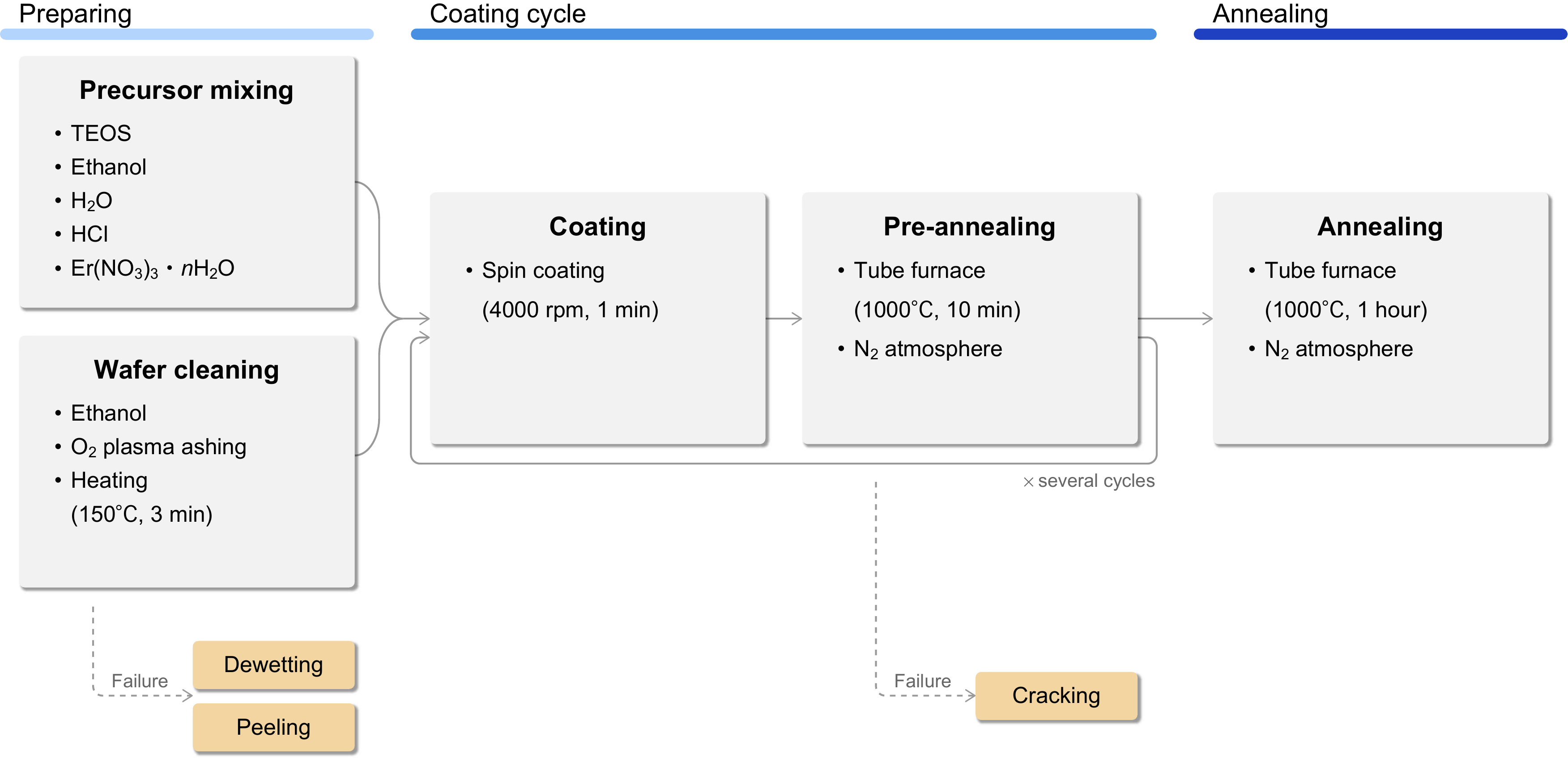}
\caption{The sol-gel thin film deposition process flow and common defects.}
\label{fig1}
\end{figure}

The sol-gel thin film process flow is shown in Fig.~\ref{fig1}. The preparation of the sol-gel solution begins with a tetraethyl orthosilicate (TEOS)~:~ethanol~:~$\mathrm{H_2O}$ ratio of 5.0~:~5.0~:~2.0. Acidic conditions are established by using hydrochloric acid to adjust the pH to around 1. Once the solution has been prepared, it is stirred for 30~min at 70$^\circ$C to facilitate the reaction and the aged at room temperature for 1~h. This aging process promotes the formation of stronger siloxane bonds, enhancing the resulting material's stability and strength. Next, the substrate is cleaned with ethanol, heated at 150$^\circ$C for 3~min, and subjected to plasma ashing. The cleaned substrate is then placed in a spin coater, and the coating is applied at 4000~rpm for 50~sec. This is followed by 10~min of pre-annealing in a nitrogen atmosphere in an annealing furnace at 1000$^\circ$C. The final step involves annealing for 1~h under the same conditions as the pre-annealing process. This process allows for the formation of a uniform sol-gel silica film with a thickness of approximately 300~nm. Repeating the coating and pre-annealing steps for thicker silica films enables the creation of films that are several \textmu m thick~\cite{Sigoli2006-aw}.

\begin{table}[htbp]
\caption{Various conditions for fabricating sol-gel thin film \label{table1}}
    \centering
    \footnotesize
    \scalebox{0.75}{
    \begin{tabular}{l c c c c c c c}
    \hline
    Parameter &\begin{tabular}{c}L. Yang\\(2005)~\cite{Yang2005-ah}\end{tabular} &\begin{tabular}{c}L. Yang\\(2005)~\cite{Yang2005-li}\end{tabular} &\begin{tabular}{c}F. Sigoli\\(2006)~\cite{Sigoli2006-aw}\end{tabular} &\begin{tabular}{c}H. Hsu\\(2009)~\cite{Hsu2009-dy}\end{tabular} &\begin{tabular}{c}A. Maker\\(2012)~\cite{Maker2012-ww}\end{tabular} & \begin{tabular}{c}H. Choi\\(2018)~\cite{Choi2018-tj}\end{tabular} & This work\\
    \hline
    \begin{tabular}{l}Water/TEOS\\(MTES)\\molar ratio\end{tabular} & 1-2               & $-$                   & $-$                   & 2                     & $-$                   & $-$                   & 4.6\\
    \hline
    \begin{tabular}{l}Stirring time\end{tabular}                   & 3 h               & 1 h                   & 24 h                  & 2 h                   & $-$                   & \begin{tabular}{c}2~h,\\24~h (aging)\end{tabular} & \begin{tabular}{c}30~min,\\1~h (aging)\end{tabular}\\
    \hline
    \begin{tabular}{l}Cosolvent\end{tabular}                       & 2-propanol        & Ethanol               & Ethanol, PVA          & Ethanol               & Ethanol               & Ethanol               & Ethanol\rule[-3mm]{0mm}{8mm}\\
    \hline
    \begin{tabular}{l}Annealing\\temperature\end{tabular}  & 1000$^\circ$C     & 500-800$^\circ$C      & 600$^\circ$C, 1200$^\circ$C & 1000$^\circ$C   & 1000$^\circ$C         & 1000$^\circ$C         & 1000$^\circ$C\\
    \hline
    \begin{tabular}{l}Annealing\\time\end{tabular}  & 3 h               & 30 min$\mathrm{^a}$   & \begin{tabular}{c}5 min (600$^\circ$C)\\1 min (1200$^\circ$C)$\mathrm{^b}$ \end{tabular} & 3 h   & 1 h  & 1 h$\mathrm{^c}$  & 1 h\\
    \hline
    \begin{tabular}{l}Film\\thickness\end{tabular}  & 1 \textmu m       & $-$                   & \begin{tabular}{c}13.50 \textmu m (600$^\circ$C)\\4.30 \textmu m (1200$^\circ$C) \end{tabular} & 1.2 \textmu m   & 0.35 \textmu m  & $-$  & 1.8 \textmu m\\
    \hline
    \multicolumn{5}{l}{$\mathrm{^a}$ 60$^\circ$C drying (5 min) before annealing}\\
    \multicolumn{5}{l}{$\mathrm{^b}$ 600$^\circ$C annealing after each coating \& 1200$^\circ$C annealing after all coatings}\\
    \multicolumn{5}{l}{$\mathrm{^c}$ 75$^\circ$C drying (5 min) before annealing}
    \end{tabular}
    }
\end{table}

However, producing defect-free films is not guaranteed under the described parameters, as the fabrication environment also plays a critical role. Adjustments must be made to account for ambient temperature, humidity, and variations in equipment, such as the stirrers and annealing furnaces used in the fabrication process. Achieving uniform sol-gel films requires careful optimization of parameters such as solvent type, stirring time, and annealing conditions, all of which can vary considerably. Therefore, it is highly useful to understand troubleshooting strategies for addressing defects that may arise.

Figure~\ref{fig1} illustrates the types of defects that can occur in sol-gel films at various stages of the fabrication process. We found that these defects can be broadly categorized into three distinct types. In the following subsections, we will discuss the causes and potential solutions for each type of defect.

\subsection{De-wetting pattern}
\begin{figure}[htbp]
\centering
\includegraphics[width=\linewidth]{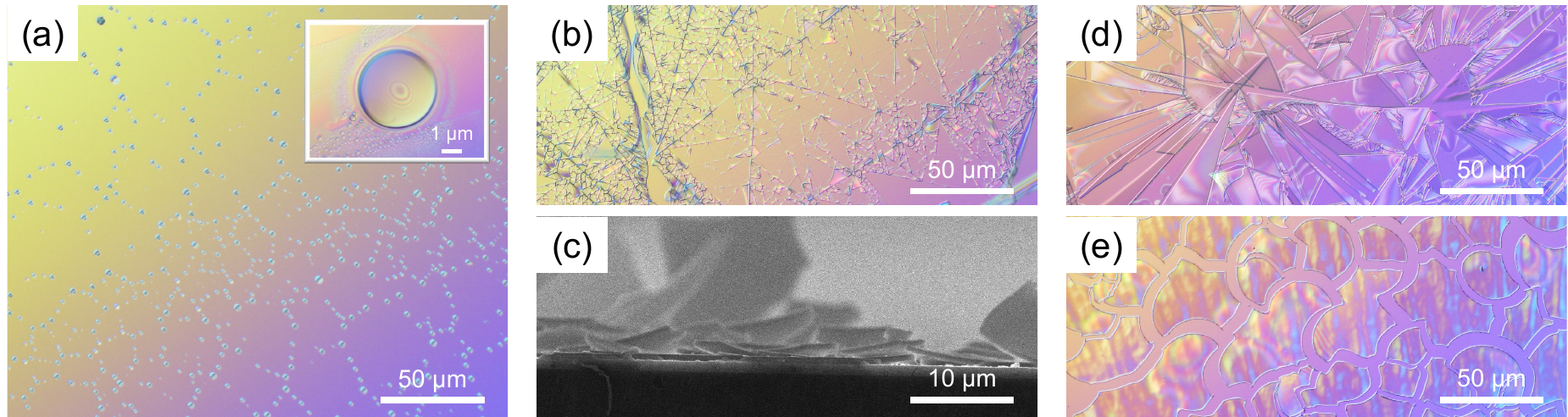}
\caption{Defects observed during sol-gel silica film deposition. Each image shows a different defect pattern: (a)~a microscopic image of a de-wetting pattern (with the inset showing a magnified image), (b)~a microscopic image of a peeling pattern, (c)~a cross-sectional SEM image of a peeling pattern, (d)~and~(e)~microscopic images of two types of cracking pattern. These patterns represent common issues encountered in thin film fabrication, resulting from different underlying causes.}
\label{fig2}
\end{figure}

In the context of sol-gel films on substrates, an agglomeration state resembling raised droplets is referred to as a de-wetting condition (Fig.~\ref{fig2}(a)), and it typically forms a hemispherical shape~\cite{Mukherjee2015-iv}. This defect has two primary causes. The first is an unsuitable substrate surface condition. If the substrate surface is not adequately cleaned and remains hydrophobic, the surface tension of the solution causes it to form droplet-like shapes. The second cause is the incomplete reaction of the solution. When the reaction time is too short, the polycondensation reaction remains incomplete, preventing the formation of a proper silica network.

If de-wetting occurs, the most effective solution is to improve the substrate surface treatment. An essential step is to perform sufficient plasma ashing (e.g., 1~min) before spin-coating the sol-gel solution, which removes organic contaminants from the substrate surface and increases its hydrophilicity. Additionally, the reaction time of the initial solution should be carefully adjusted.

\subsection{Peeling pattern}
As shown in Fig.~\ref{fig2}(b), thin, mesh-like peeling can occur. A cross-sectional SEM image of this substrate shown in Fig.~\ref{fig2}(c). This phenomenon arises when the layers do not bond properly with each other. A poorly hydrophilic substrate surface is a primary cause of inadequate bonding between sol-gel layers. Additionally, while substances evaporated during annealing are generally removed, residual organic materials that are not fully eliminated can remain on the surface~\cite{Idris2018-ei}. These residues interfere with the formation of vertical silica networks, leading to the observed peeling pattern.

A practical solution to this issue is to perform $\mathrm{O_2}$ plasma ashing on the substrate surface before each layer is spin coated. Plasma ashing effectively removes organic residues, ensuring better adhesion between layers. When using this method, we observed no delamination, even when stacking multiple layers of films.

\subsection{Cracking pattern}
Among the observed defects, the most common is a pattern of cracks, as seen in the microscopic images in Figs.~\ref{fig2}(d) and \ref{fig2}(e). These cracks are caused by excessive film thickness after spin coating, which leads to defects during annealing. As the sol-gel solution is spin-coated and then annealed, the solvent evaporates from the film, resulting in densification of the silica network and a reduction in film volume. If the annealing temperature is too low, this volume reduction induces structural instability, which results in cracks or delamination. Conversely, structural relaxation is more likely at high annealing temperatures of around 1000$^\circ$C, which promotes uniform film formation. Therefore, it is preferable to place the coated substrate in a preheated annealing furnace rather than gradually increasing the temperature. Additionally, since the glass transition temperature of sol-gel films is approximately 600$^\circ$C~\cite{Innocenzi1994-vr}, the annealing temperature should exceed this threshold. Volume shrinkage during annealing can reach up to 20$\mathrm{\%}$, so each coating layer should be thin enough to allow structural relaxation to compensate for shrinkage. In this study, each layer is approximately 300~nm thick.

In Fig.~\ref{fig2}(d), although the mixing ratio of the starting solution is optimal, an extended reaction time of 3~h results in an excessively thick film, which induces stress and contributes to crack formation. An uneven film thickness or significant variations during spin coating can also cause these patterns. This unevenness arises from differences between the centrifugal forces at the center and the edges of the substrate during spin coating, which affect the solvent evaporation rate from the sol-gel solution. Evaporation begins as soon as the solution is dropped onto the silicon substrate and intensifies as the substrate spins, leading to a thickness difference of about 150~nm between the center and edges. Therefore, careful control of the solution’s reaction time and spin-coating conditions is essential.

\begin{figure}[htbp]
\centering
\includegraphics[width=\linewidth]{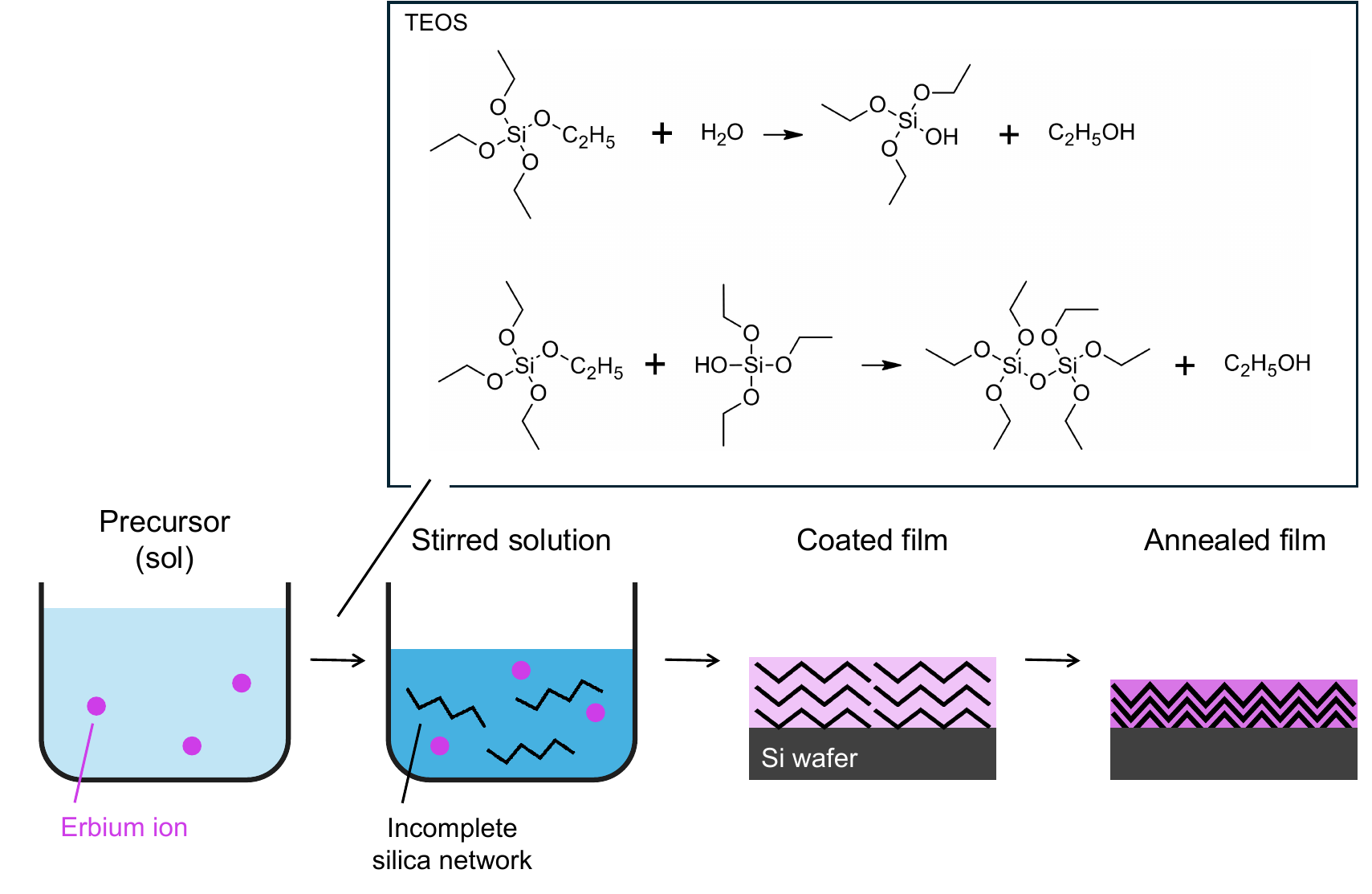}
\caption{Schematic illustration of the sol-gel reaction. In the precursor containing erbium ions, hydrolysis and polycondensation reactions occur over the stirring time, leading to the formation of an incomplete silica network. After coating onto the substrate, organic solvents and water are removed during the annealing process, resulting in a denser silica thin film.}
\label{fig3}
\end{figure}

In Fig.~\ref{fig2}(e), we observe patterns resembling “jigsaw puzzles” which we attribute to differences in water content in the starting solutions~\cite{Slooff2001-bs}. For this sample, the molar ratio of $\mathrm{H_2O}$ to TEOS in the starting solution is 3.0. According to the reaction formula (Fig.~\ref{fig3}) and reaction efficiency, more than 4 moles of water per mole of TEOS are needed for a proper reaction. The cracks in this sample can thus be attributed to insufficient water content. This deficiency hinders the formation of the silica network, which is essential for structural integrity, as insufficient water limits the formation of the siloxane bonds (Si–O bonds) necessary for a robust silica network. Consequently, island-like cracks form, with larger island sizes observed in solutions with higher water content.

\section{Film evaluation and microresonator fabrication}
\begin{figure}[htbp]
\centering
\includegraphics[width=\linewidth]{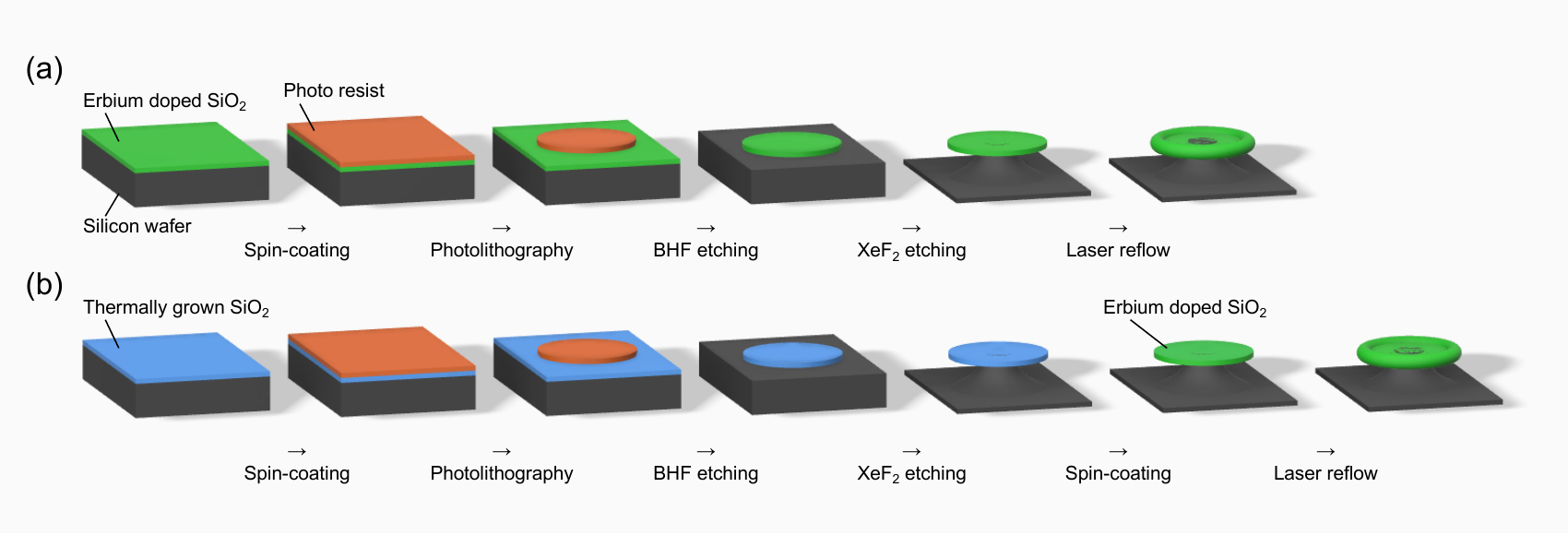}
\caption{(a) Schematic illustration of the process flow for fabricating a rare-earth ion-doped silica toroid from a thin sol-gel film. (b) Alternative fabrication method for a rare-earth ion-doped silica toroid by coating the surface of a pre-fabricated toroid with a sol-gel film.}
\label{fig4}
\end{figure}

We fabricated an erbium-doped microtoroid after confirming the high quality of the defect-free ion-doped sol-gel film deposited on a silicon wafer. The fabrication process, illustrated in Fig.~\ref{fig4}(a), follows a conventional photolithographic approach. The microtoroid has a diameter of 60~\textmu m and an erbium ion concentration of $\mathrm{2.0\times10^{-19}~cm^{-3}}$, comparable to commercial erbium-doped fibers. Additionally, when a microtoroid is fabricated using a sol-gel thin film without erbium ion doping, it exhibits a $Q$-factor of approximately $10^7$, similar to microtoroids made from thermally oxidized films~\cite{Armani2003-ee}.

It is important to note that, due to the limited thickness of sol-gel thin films, the fabrication method reported in previous studies is suitable for resonators with diameters of about 100~\textmu m or less. The method described in the next section should be employed for larger toroid resonators.

\section{Coating the cavity for large-size Er-doped microtoroid fabrication}
\subsection{Fabrication}
Next, we coated microstructures with sol-gel films. As shown in Fig.~\ref{fig4}(b), the process flow involves coating a microdisk resonator with a sol-gel thin film and reflow using a $\mathrm{CO_2}$ laser. When fabricating state-of-the-art silica disks or silica toroid resonators, increasing the resonator diameter often results in buckling due to the mismatch between the thermal expansion coefficients of the silica film and the silicon substrate~\cite{Chen2013-ra,Ma2017-ar}. To prevent buckling, it is known that the silica film thickness must exceed 5~\textmu m. However, since each sol-gel thin film layer is limited to approximately 300~nm per coating, fabricating such a resonator using the method in Fig.~\ref{fig4}(a) presents a challenge.

\begin{figure}[htbp]
\centering
\includegraphics[width=\linewidth]{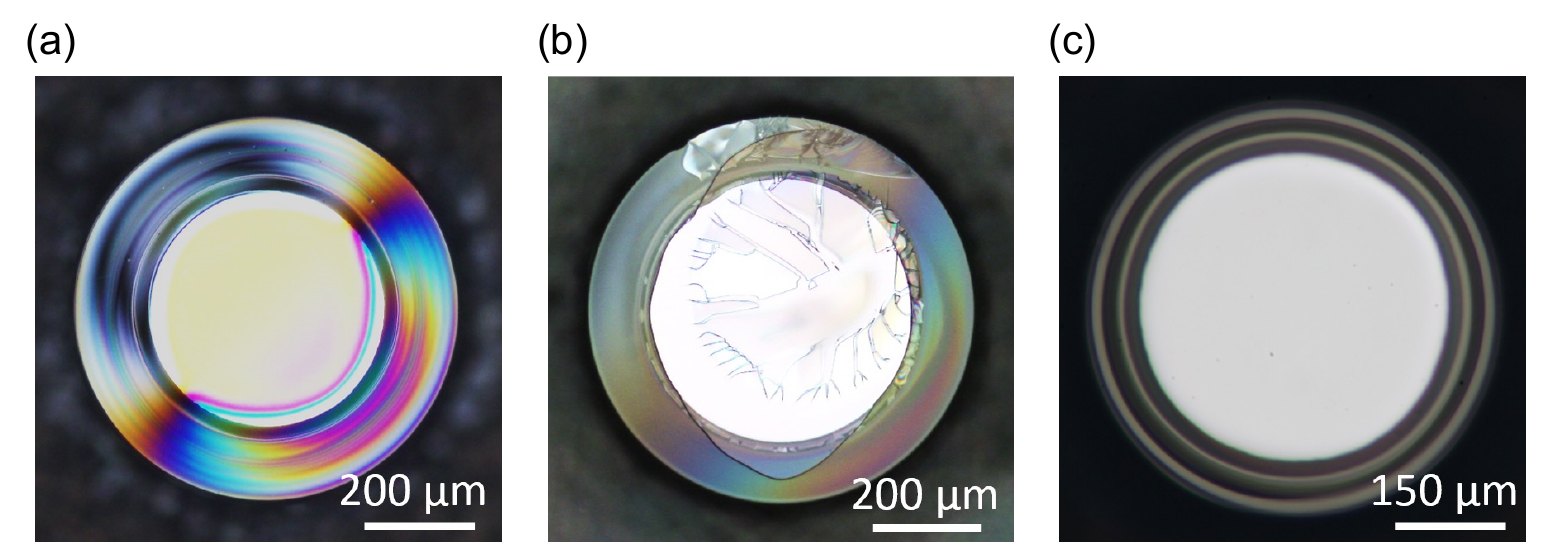}
\caption{(a) Silica micro disc coated with sol-gel film. (b) Silica micro disc coated with sol-gel film but without ethanol atmosphere. (c) After reflowing (a).}
\label{fig5}
\end{figure}

On the other hand, direct coating onto the silica disk resonator can circumvent the issue of buckling. In this experiment, we fabricated disk resonators with a diameter of 600~\textmu m using an 8~\textmu m thick thermally oxidized silica film, and no buckling was observed in these resonators. For coating, we used a spin coater set at 4000~rpm for 45~s, followed by annealing at 1000$^\circ$C for 1~h. The fabricated device is shown in Fig.~\ref{fig5}(a). The $Q$-value of the disk resonators before and after coating remained high at $1.2\times10^6$, indicating that the coating has no impact on the resonator’s performance. This coating method has significant potential for various applications, as it allows for ion activation without altering the resonator's quality.

Defects, as shown in Fig.~\ref{fig5}(b), can occur during spin coating; however, coating without cracks is achievable by establishing an ethanol atmosphere inside the spin coater~\cite{Syms1994-ld,Min1995-op}.

After successfully obtaining the sol-gel activated silica disk, we employed $\mathrm{CO_2}$ laser reflow, reducing the diameter to approximately 450~\textmu m (Fig.~\ref{fig5}(c)), as estimated from the measured free spectral range.

\subsection{Optical measurement of the sol-gel coated microtoroid}
\begin{figure}[htbp]
\centering
\includegraphics[width=\linewidth]{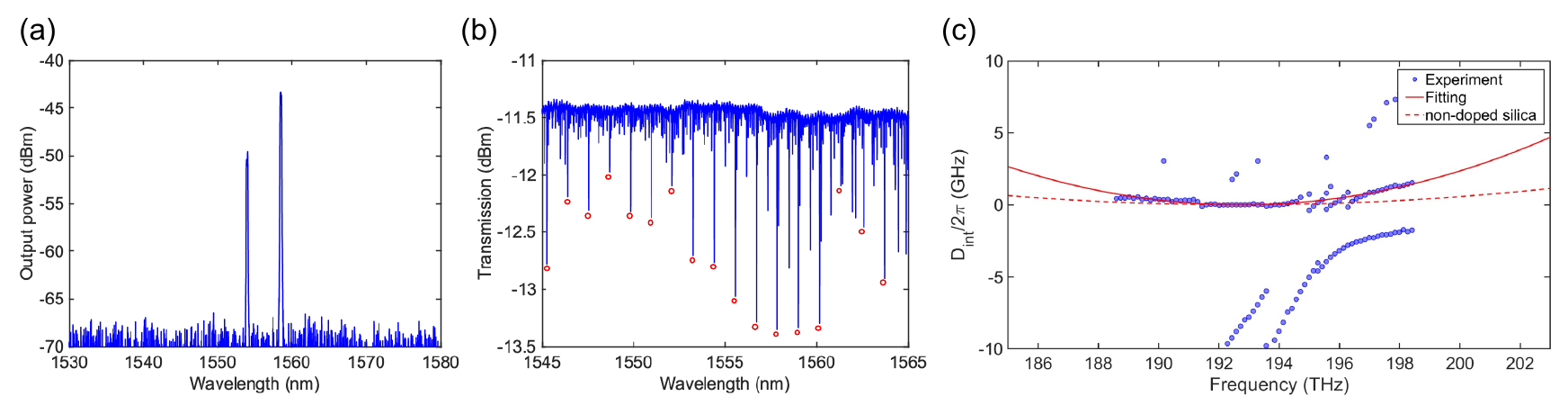}
\caption{(a) Laser emission spectrum of an erbium-doped microtoroid fabricated using the coating method. When pumped in the 1480 nm~band, lasing occurs in the 1550~nm band, with an efficiency of 1.5$\mathrm{\%}$. (b) The transmission spectrum of the erbium-doped microtoroid in the 1550~nm band, with red circles indicating the fundamental TE mode. (c) Dispersion measurement results for the fundamental TE mode: blue dots represent the measured values, and the solid red line is the fitted curve. The dashed red line represents the theoretical dispersion values for a non-doped silica microtoroid with the same structure. Adding erbium ions has minimal impact on the dispersion.}
\label{fig6}
\end{figure}

The erbium-doped toroidal resonator with a diameter of 450~\textmu m, shown in Fig.~\ref{fig5}(c), exhibits a $Q$-value of $3.5\times10^6$ in the 1480~nm excitation mode. As the pump power is increased to excite the erbium ions, green up-conversion light is observed, indicating a good overlap between the erbium ions and the optical mode. When the pump power exceeds 350~\textmu W, multi-mode laser oscillations are observed in the 1550~nm band, as shown in Fig.~\ref{fig6}(a). Thus, the lasing threshold is approximately 350~\textmu W, with a lasing efficiency of 1.5$\mathrm{\%}$.

The transmission spectrum in the 1550~nm band for this cavity is shown in Fig.~\ref{fig6}(b), with the red circles indicating the TE modes. We measured the dispersion of this cavity, as shown in Fig.~\ref{fig6}(c)~\cite{Fujii2020-gg}. The blue dots represent the measured values, while the solid red line shows the fitting curve, which accounts for up to third-order dispersion. The dashed red line represents the theoretical dispersion curve for a silica microtoroid, assuming no erbium ion doping and identical major and minor diameters. The measurement results indicate that erbium ion doping has minimal impact on the dispersion, allowing the resonator to maintain anomalous dispersion.

\section{Summary}
In this paper, we highlighted the sol-gel method as an efficient and cost-effective technique that does not require large-scale equipment, making it an attractive option for producing small laser devices. However, previous studies lacked standardized fabrication conditions for sol-gel thin films. To address this, we provided detailed insights into common thin film defects, their underlying causes, and potential solutions. Our goal was to establish a smooth troubleshooting process and a reproducible method for sol-gel film fabrication, enabling researchers to consistently achieve high-quality films.

Using the sol-gel coating method, we successfully fabricated erbium-doped microtoroid resonators, overcoming the previous diameter limitation of resonators of below 100~\textmu m. Our approach allowed the creation of buckling-free resonators with diameters as large as 450~\textmu m. The erbium-doped microtoroid resonators demonstrated laser oscillation in the 1550~nm band.

This study provides a reliable framework for fabricating erbium-doped optical devices, such as gain-doped photonic integrated circuits. It also demonstrates the potential of the sol-gel method for producing high-performance devices.

\begin{backmatter}
\bmsection{Funding}
This work was supported by MEXT Quantum Leap Flagship Program (MEXT Q-LEAP) JPMXS0118067246 and by JSPS KAKENHI (JP24K17624). S. Fujii acknowledges the support from Inamori Foundation and Nippon Sheet Glass Foundation for Materials Science and Engineering.


\bmsection{Acknowledgment}
The authors thank Mr. H. Kumazaki for providing technical support fot this work.

\bmsection{Data availability} Data underlying the results presented in this paper are not publicly available at this time but may be obtained from the authors upon reasonable request.

\end{backmatter}

\bibliography{Solgel_ref}

\begin{thebibliography}{10}
\newcommand{\enquote}[1]{``#1''}

\bibitem{Polman1997-va}
A.~Polman, \enquote{Erbium implanted thin film photonic materials,} {\protect\JournalTitle{Journal of applied physics}} \textbf{82}, 1--39 (1997).

\bibitem{Bradley2011-rg}
J.~D.~B. Bradley and M.~Pollnau, \enquote{Erbium-doped integrated waveguide amplifiers and lasers,} {\protect\JournalTitle{Laser \& photonics reviews}} \textbf{5}, 368--403 (2011).

\bibitem{Mu2020-nf}
J.~Mu, M.~Dijkstra, J.~Korterik, \emph{et~al.}, \enquote{High-gain waveguide amplifiers in {Si}$_{3}${N}$_{4}$ technology via double-layer monolithic integration,} {\protect\JournalTitle{Photonics research}} \textbf{8}, 1634 (2020).

\bibitem{Xin2019-sq}
M.~Xin, N.~Li, N.~Singh, \emph{et~al.}, \enquote{Optical frequency synthesizer with an integrated erbium tunable laser,} {\protect\JournalTitle{Light, science \& applications}} \textbf{8}, 122 (2019).

\bibitem{Liu2022-ld}
Y.~Liu, Z.~Qiu, X.~Ji, \emph{et~al.}, \enquote{A photonic integrated circuit-based erbium-doped amplifier,} {\protect\JournalTitle{Science}} \textbf{376}, 1309--1313 (2022).

\bibitem{Liu2024-qx}
Y.~Liu, Z.~Qiu, X.~Ji, \emph{et~al.}, \enquote{A fully hybrid integrated erbium-based laser,} {\protect\JournalTitle{Nature photonics}} \textbf{18}, 829--835 (2024).

\bibitem{Prugger_Suzuki2021-wi}
T.~S. Prugger~Suzuki, A.~Nakashima, K.~Nagashima, \emph{et~al.}, \enquote{Design of a passively mode-locking whispering-gallery-mode microlaser,} {\protect\JournalTitle{JOSA B}} \textbf{38}, 3172--3178 (2021).

\bibitem{Kumagai2018-eg}
T.~Kumagai, N.~Hirota, K.~Sato, \emph{et~al.}, \enquote{Saturable absorption by carbon nanotubes on silica microtoroids,} {\protect\JournalTitle{Journal of applied physics}} \textbf{123}, 233104 (2018).

\bibitem{Tan2021-ib}
T.~Tan, Z.~Yuan, H.~Zhang, \emph{et~al.}, \enquote{Multispecies and individual gas molecule detection using stokes solitons in a graphene over-modal microresonator,} {\protect\JournalTitle{Nature communications}} \textbf{12}, 6716 (2021).

\bibitem{Fujii2024-bi}
S.~Fujii, N.~Fang, D.~Yamashita, \emph{et~al.}, \enquote{van der waals decoration of ultra-high-\textit{Q} silica microcavities for $\chi$$^{(2)}$--$\chi$$^{(3)}$ hybrid nonlinear photonics,} {\protect\JournalTitle{Nano letters}} \textbf{24}, 4209--4216 (2024).

\bibitem{Wang2023-bn}
C.~Wang, Z.~Fu, W.~Mao, \emph{et~al.}, \enquote{Non-hermitian optics and photonics: from classical to quantum,} {\protect\JournalTitle{Advances in optics and photonics}} \textbf{15}, 442 (2023).

\bibitem{Chen2018-lk}
W.~Chen, D.~Leykam, Y.~D. Chong, and L.~Yang, \enquote{Nonreciprocity in synthetic photonic materials with nonlinearity,} {\protect\JournalTitle{MRS bulletin}} \textbf{43}, 443--451 (2018).

\bibitem{Zhang2019-lt}
M.~Zhang, W.~Sweeney, C.~W. Hsu, \emph{et~al.}, \enquote{Quantum noise theory of exceptional point amplifying sensors,} {\protect\JournalTitle{Physical review letters}} \textbf{123} (2019).

\bibitem{Wang2021-tx}
C.~Wang, W.~R. Sweeney, A.~D. Stone, and L.~Yang, \enquote{Coherent perfect absorption at an exceptional point,} {\protect\JournalTitle{Science (New York, N.Y.)}} \textbf{373}, 1261--1265 (2021).

\bibitem{Zhang2018-js}
J.~Zhang, B.~Peng, {\c{S}}.~K. {\"{O}}zdemir, \emph{et~al.}, \enquote{A phonon laser operating at an exceptional point,} {\protect\JournalTitle{Nature photonics}} \textbf{12}, 479--484 (2018).

\bibitem{Chen2017-sk}
W.~Chen, {\c{S}}.~Kaya~{\"{O}}zdemir, G.~Zhao, \emph{et~al.}, \enquote{Exceptional points enhance sensing in an optical microcavity,} {\protect\JournalTitle{Nature}} \textbf{548}, 192--196 (2017).

\bibitem{Huang2017-bh}
Y.~Huang, Y.~Shen, C.~Min, \emph{et~al.}, \enquote{Unidirectional reflectionless light propagation at exceptional points,} {\protect\JournalTitle{Nanophotonics}} \textbf{6}, 977--996 (2017).

\bibitem{Peng2014-pc}
B.~Peng, S.~K. {\"{O}}zdemir, S.~Rotter, \emph{et~al.}, \enquote{Loss-induced suppression and revival of lasing,} {\protect\JournalTitle{Science (New York, N.Y.)}} \textbf{346}, 328--332 (2014).

\bibitem{Peng2014-ry}
B.~Peng, {\c{S}}.~K. {\"{O}}zdemir, F.~Lei, \emph{et~al.}, \enquote{Parity--time-symmetric whispering-gallery microcavities,} {\protect\JournalTitle{Nature physics}} \textbf{10}, 394--398 (2014).

\bibitem{Imamura2024-pa}
R.~Imamura, S.~Fujii, A.~Nakashima, and T.~Tanabe, \enquote{Exceptional point proximity-driven mode-locking in coupled microresonators,} {\protect\JournalTitle{Optics express}} \textbf{32}, 22280 (2024).

\bibitem{Min2004-zk}
B.~Min, T.~J. Kippenberg, L.~Yang, \emph{et~al.}, \enquote{Erbium-implanted high-$q$ silica toroidal microcavity laser on a silicon chip,} {\protect\JournalTitle{Physical review. A}} \textbf{70}, 033803 (2004).

\bibitem{Orignac1999-km}
X.~Orignac, D.~Barbier, X.~Min~Du, \emph{et~al.}, \enquote{Sol--gel silica/titania-on-silicon er/yb-doped waveguides for optical amplification at 1.5 $\mu${m},} {\protect\JournalTitle{Optical materials}} \textbf{12}, 1--18 (1999).

\bibitem{Zhang2013-sf}
X.~Zhang and A.~M. Armani, \enquote{Silica microtoroid resonator sensor with monolithically integrated waveguides,} {\protect\JournalTitle{Optics express}} \textbf{21}, 23592--23603 (2013).

\bibitem{Yang2003-zt}
L.~Yang, D.~K. Armani, and K.~J. Vahala, \enquote{Fiber-coupled erbium microlasers on a chip,} {\protect\JournalTitle{Applied physics letters}} \textbf{83}, 825--826 (2003).

\bibitem{Yang2005-ah}
L.~Yang, T.~Carmon, B.~Min, \emph{et~al.}, \enquote{Erbium-doped and raman microlasers on a silicon chip fabricated by the sol--gel process,} {\protect\JournalTitle{Applied physics letters}} \textbf{86}, 091114 (2005).

\bibitem{Liu2020-zp}
S.~Liu, R.~Lv, Y.~Wang, \emph{et~al.}, \enquote{Passively mode-locked fiber laser with {WS2}/{SiO2} saturable absorber fabricated by sol-gel technique,} {\protect\JournalTitle{ACS applied materials \& interfaces}} \textbf{12}, 29625--29630 (2020).

\bibitem{Huang2002-jq}
W.~Huang, R.~R.~A. Syms, E.~M. Yeatman, \emph{et~al.}, \enquote{Fiber-device-fiber gain from a sol-gel erbium-doped waveguide amplifier,} {\protect\JournalTitle{IEEE photonics technology letters: a publication of the IEEE Laser and Electro-optics Society}} \textbf{14}, 959--961 (2002).

\bibitem{Kozuka2004-kp}
H.~Kozuka and M.~Komeda, \enquote{Effect of the amount of water for hydrolysis on cracking and stress evolution in alkoxide-derived sol-gel silica coating films,} {\protect\JournalTitle{Journal of the Ceramic Society of Japan, Supplement}} \textbf{112}, S223--S227 (2004).

\bibitem{Yahata2009-td}
R.~Yahata and H.~Kozuka, \enquote{Stress evolution of sol--gel-derived silica coatings during heating: The effects of the chain length of alcohols as solvents,} {\protect\JournalTitle{Thin solid films}} \textbf{517}, 1983--1988 (2009).

\bibitem{Sigoli2006-aw}
F.~A. Sigoli, R.~R. Gon\c{c}alves, Y.~Messaddeq, and S.~J.~L. Ribeiro, \enquote{Erbium- and ytterbium-doped sol--gel {SiO2}--{HfO2} crack-free thick films onto silica on silicon substrate,} {\protect\JournalTitle{Journal of Non-Crystalline Solids}} \textbf{352}, 3463--3468 (2006).

\bibitem{Yang2005-li}
L.-L. Yang, Y.-S. Lai, J.-S. Chen, \emph{et~al.}, \enquote{Compositional tailored sol-gel {SiO2}-{TiO2} thin films: Crystallization, chemical bonding configuration, and optical properties,} {\protect\JournalTitle{Journal of materials research}} \textbf{20}, 3141--3149 (2005).

\bibitem{Hsu2009-dy}
H.-S. Hsu, C.~Cai, and A.~M. Armani, \enquote{Ultra-low-threshold er:yb sol-gel microlaser on silicon,} {\protect\JournalTitle{Optics express}} \textbf{17}, 23265--23271 (2009).

\bibitem{Maker2012-ww}
A.~J. Maker, B.~A. Rose, and A.~M. Armani, \enquote{Tailoring the behavior of optical microcavities with high refractive index sol-gel coatings,} {\protect\JournalTitle{Optics letters}} \textbf{37}, 2844--2846 (2012).

\bibitem{Choi2018-tj}
H.~Choi and A.~M. Armani, \enquote{Raman--kerr frequency combs in zr-doped silica hybrid microresonators,} {\protect\JournalTitle{Optics Letters}} \textbf{43}, 2949--2952 (2018).

\bibitem{Mukherjee2015-iv}
R.~Mukherjee and A.~Sharma, \enquote{Instability{,} self-organization and pattern formation in thin soft films,} {\protect\JournalTitle{Soft matter}} \textbf{11}, 8717--8740 (2015).

\bibitem{Idris2018-ei}
A.~S. Idris, H.~Jiang, and K.~Hamamoto, \enquote{Multi-layer stacking scheme of sol-gel based {SiO}$_{2}$ towards thicker (>0.8 $\mathrm{\mu}${m}) cladding layers for optical waveguides,} {\protect\JournalTitle{IEICE Electronics Express}} \textbf{15}, 20180783--20180783 (2018).

\bibitem{Innocenzi1994-vr}
P.~Innocenzi, M.~O. Abdirashid, and M.~Guglielmi, \enquote{Structure and properties of sol-gel coatings from methyltriethoxysilane and tetraethoxysilane,} {\protect\JournalTitle{Journal of Sol-Gel Science and Technology}} \textbf{3}, 47--55 (1994).

\bibitem{Slooff2001-bs}
L.~H. Slooff, M.~J.~A. de~Dood, A.~van Blaaderen, and A.~Polman, \enquote{Effects of heat treatment and concentration on the luminescence properties of erbium-doped silica sol--gel films,} {\protect\JournalTitle{Journal of Non-Crystalline Solids}} \textbf{296}, 158--164 (2001).

\bibitem{Armani2003-ee}
D.~K. Armani, T.~J. Kippenberg, S.~M. Spillane, and K.~J. Vahala, \enquote{Ultra-high-{Q} toroid microcavity on a chip,} {\protect\JournalTitle{Nature}} \textbf{421}, 925--928 (2003).

\bibitem{Chen2013-ra}
T.~Chen, H.~Lee, and K.~J. Vahala, \enquote{Thermal stress in silica-on-silicon disk resonators,} {\protect\JournalTitle{Applied physics letters}} \textbf{102}, 031113 (2013).

\bibitem{Ma2017-ar}
J.~Ma, X.~Jiang, and M.~Xiao, \enquote{Kerr frequency combs in large-size, ultra-high-{Q} toroid microcavities with low repetition rates [invited],} {\protect\JournalTitle{Photonics Research}} \textbf{5}, B54--B58 (2017).

\bibitem{Syms1994-ld}
R.~R.~A. Syms and A.~S. Holmes, \enquote{Deposition of thick silica-titania sol-gel films on si substrates,}  (1994).

\bibitem{Min1995-op}
X.~Min, X.~Orignac, and R.~M. Almeida, \enquote{Striation-free, spin-coated sol-gel optical films,}  (1995).

\bibitem{Fujii2020-gg}
S.~Fujii and T.~Tanabe, \enquote{Dispersion engineering and measurement of whispering gallery mode microresonator for kerr frequency comb generation,} {\protect\JournalTitle{Nanophotonics}} \textbf{9}, 1087--1104 (2020).

\end{thebibliography}






\end{document}